\documentclass[
aps,nofootinbib,
showpacs,showkeys,preprint
tightenlines,preprintnumbers,] {revtex4}

\usepackage{epsf,epsfig,subfigure,graphicx,amsmath,amssymb 
}
\usepackage{color}
\newcommand{\dis}[1]{\begin{equation}\begin{split}#1\end{split}\end{equation}}
\newcommand{\ie}{{\it i.e.~}}
\newcommand{\etal}{{\it et al.\,}}

\newcommand{\Qem}{Q_{\rm em}}

\newcommand{\meV}{\,\mathrm{MeV}}

\def\sw0{{$\sin^2\theta_W^0$}}

\newcommand{\Z}{{\bf Z}}

\def\smg{{SU(3)$_C\times$SU(2)$_W\times$U(1)$_Y$}}
\def\E6{{\rm E_6}}

\def\EE8{{\rm E_8\times E_8'}}

\def\three{{\bf 3}}

\def\one{{\bf 1}}
\def\oneb{{\bar{\bf 1}}}
\def\two{{\bf 2}}
\def\five{{\bf 5}}
\def\ten{{\bf 10}}
\def\tenb{{\overline{\bf 10}}}

\def\fiveb{{\overline{\bf 5}}}

\def\three{{\bf 3}}

\begin{document}

\draft

\title{\Large\bf  Tetrahedral  A$_{\bf 4}$ Symmetry in Anti-SU(5) GUT
}

\author{ Paul H. Frampton$^{(1)}$, 
Jihn E.  Kim$^{(2,3)}$, Se-Jin Kim$^{(2)}$, Soonkeon Nam$^{(2)}$ }

\address{$^{(1)}$ Dipartimento di Matematica e Fisica ``Ennio De Giorgi'', Universita del Salento and INFN-Lecce,
Via Arnesano,  73100 Lecce, Italy,  \\
$^{(2)}$Department of Physics, Kyung Hee University, 26 Gyungheedaero, Dongdaemun-Gu, Seoul 02447, Republic of Korea,  \\
 $^{(3)}$ 
Department of Physics and Astronomy, Seoul National University, 1 Gwanakro, Gwanak-Gu, Seoul 08826, Republic of Korea
}

\begin{abstract} 
\noindent
We  construct a flavor model in an anti-SU(5) GUT with a tetrahedral symmetry $A_4$.  We choose a basis where $\Qem=-\frac13$ quarks and charged leptons are already mass eigenstates.  This choice is possible from the $A_4$ symmetry. Then, matter representation $\tenb_{-1}^{\rm\, matter}$ contains both a quark doublet and a heavy neutrino $N$, which enables us to use the $A_4$ symmetry to both  $\Qem=+\frac23$ quark masses and neutrino masses (through the see-saw via $N$). This is made possible because the anti-SU(5) breaking is achieved by the Higgs fields transforming as anti-symmetric representations of SU(5), $\tenb_{-1}^H\oplus \ten_{+1}^H$, reducing the rank-5 anti-SU(5) group down to the rank-4 standard model group \smg. For possible mass matrices, the $A_4$ symmetry predictions on mass matrices at field theory level are derived. Finally, an illustration from string compactification is presented.

\end{abstract}
\maketitle

\section{Introduction}\label{sec:Introduction}

\bigskip
\noindent
Recently, we pointed out analytically how the tetrahedral  discrete symmetry $A_4$ results from the permutation symmetry $S_4$ \cite{FK19}. The $A_4$ discrete symmetry \cite{Ma01,Ma02,Ma04,Altarelli05,Lam07,Morisi07,
Morisi08, Blum08,Altarelli09,Zee13} in connection with the tri-bimaximal form of the Pontecorvo-Maki-Nakagawa-Sakata (PMNS) lepton mixing matrix \cite{PMNS1, PMNS2, PMNS3} has been observed long time ago. The underlying permutation symmetry is useful in model building and furthermore it can be accomodated to string compactification. In string compactification, chiral fields can arise from fixed points also \cite{KimRp19}. The multiplicity $N$ in a fixed point  should respect permutation symmetry $S_N$ because the chiral fields at that fixed point are not distinguished. In this paper,  we will  use a specific grand unified theory (GUT) anti-SU(5) \cite{Barr82,DKN84}.

\bigskip
\noindent
Georgi and Glashow's GUT SU(5) \cite{GG74} is an important prototype in the consideration of GUTs. An initial success was attributed to the $b-\tau$ unification \cite{Buras78}. However, there may be two issues against the GG model when one tries to include it in an  ultraviolet completed theory. The rank of the GG group is 4 which is identical to that of the Standard Model (SM) gauge group \smg. Therefore, string compactification, an ultraviolet completion of the GG SU(5), needs an adjoint representation for breaking the GG SU(5) down to the SM gauge group without changing the rank. Firstly, in string compactification, it is not possible to obtain an adjoint representation at the level-1 construction \cite{KSChoiBk}.
Second, the $\Qem=-\frac13$ Georgi--Jarlskog quark mass relations \cite{GeorgiJarlskog} need another  representation {\bf 45} beyond a quintet of Higgs fields. The need for this additional representation makes it difficult for it to be  realizesed in the string compactification. Of course, one may argue that {\bf 45} may arise from non-renormalisable interactions, which needs another fine-tuning.

\bigskip
\noindent
Therefore, the anti-SU(5) or flipped SU(5) is preferred in string compactification. Barr commented that flipped-SU(5) is a subgroup of SO(10) \cite{Barr82} but here we consider it an independent GUT since  string compactification may not go through an intermediate SO(10) which also needs an adjoint representation for spontaneous symmetry breaking to obtain Barr's flipped SU(5). On the other hand,   For breaking anti-SU(5), we use a vectorlike representation $\ten_{-1}$ and $\tenb_{+1}$ (the subscripts are $X$ charges) which are anti-symmetric tensor representations of SU(5) and hence it is called `anti-SU(5)' in \cite{DKN84}. This generalization for spontaneous symmetry breaking by anti-symmetric representations in string compactification stops at SU(7) \cite{KimSU7}. 

\bigskip
\noindent
Since the anti-SU(5)   gauge group  SU(5)$\times$U(1)$_X$ is rank-5, one can use anti-symmetric represenations to reduce rank 1 to arrive at the rank-4 SM gauge group via the Higgs fields,
\dis{
\tenb_{-1}^H=\Big\{(\three,\one)^c_L, (\three,\two)_{L}, B^{45} \Big\}_{-1}  ,~\ten_{+1}^H=\Big\{(\three,\one)_L, (\three,\two)^c_{L}, B_{45} \Big\}_{+1},\label{eq:GUTHiggs}
}
we use the $X$ definition given in Ref. \cite{KimRp19}.
 The vacuum expectation values (VEVs) of neutral singlets $ B^{45}$ and  $ B_{45}$ (in $\tenb_{-1}^H$ and  $\ten_{+1}^H$) break the anti-SU(5) down to the SM gauge group. But, there is no $b-\tau$ unification in this anti-SU(5).

\bigskip
\noindent
One family in the anti-SU(5) in terms of   left-handed (L-handed) fields is
\dis{
\tenb_{-1}=\Big\{(d^\alpha)^c_L, {Q}^{\alpha }_{L}, N^c_L \Big\}  ,~\five_{+3}= \Big\{(u^\alpha)^c_L,\ell_{L} \Big\}  ,~\oneb_{-5}= e^c_L,\label{eq:DefF5}
}
where
\dis{
{Q}^{\alpha}_{L} =\begin{pmatrix} u^{\alpha}\\[0.3em] d^{\alpha }\end{pmatrix}_L,~{\ell}_{L} =\begin{pmatrix} \nu_e \\[0.3em] e\end{pmatrix}_L.
}
Note that all SU(2) singlets are with superscript $^c$. So, the singlet neutrino $N^c_L$ is in $\bar{\tenb}_{-1}$ and   $N_R$ has $X=+1$.
To break the SM gauge group to U(1)$_{\rm em}$, we need a Higgs quintet(s) $\fiveb^H_{+2}$ and $\five^H_{-2}$. 

\bigskip

\noindent
The family problem or the flavor problem consists of two parts. Firstly, why are there three families which have exactly the same gauge interactions. Second, why do these families  have different  Yukawa couplings? In GUTs, the first problem was formulated by Georgi \cite{Georgi79} which was applied in extended GUTs \cite{Kim80,Frampton79}. In string theory, three family models have been searched in various compactification schemes \cite{Candelas,Dixon2,Ibanez1,Tye87,
Bachas87,Gepner87,IKNQ,Munoz88,Lykken96,PokorskiW99,
Cleaver99,Cleaver01,CleaverNPB,Donagi02,Raby05,
Donagi05,He05,Donagi06,He06,Blumenhagen06,Cvetic06,
Blumenhagen07,KimJH07,Faraggi07,Cleaver07,Munoz07,
Nilles08}. The second problem is usually talked in terms of flavor symmetry. The flavor symmetry is designed to calculate the CKM and PMNS matrices. Permutation symmetry $S_3$ has been started to calculate the CKM matrix \cite{Pakvasa78,Segre79} but permutation symmetries blossomed recently in fitting the PMNS matrix \cite{PDG18PMNS}.

\bigskip
\noindent
In Sec. \ref{sec:FramptonKim}, we summarize the results of Ref. \cite{FK19}. In Sec. \ref{sec:Yukawa}, we discuss the $A_4$ symmetry at field theory level for three families in the anti-SU(5) GUT. We obtain possible forms of mass matrices of quarks and leptons, which are related by the anti-SU(5) representations. In Sec. \ref{sec:Z12Rp}, we present an example for possible quark and lepton mass matrices in a string derived spectra presented in Ref. \cite{KimRp19}. Finally, a brief conclusion is given in Sec. \ref{sec:Conclusion}.

\section{$A_4$ from $S_4$}\label{sec:FramptonKim}

\noindent
The permutation symmetry $S_3$ has been used in the leptonic sector for a bimaximal PMNS matrix in the late 1990s \cite{Perkins95,Kang97}, and the $A_4$ symmetry has been started in the early 2000s \cite{Ma02}. 
 The flavor symmetry in  the PMNS matrix of a tri-bimaximal  form
 \dis{
V\sim  \begin{pmatrix}
 \times &\times& \times\\[0.5em]
\frac{1}{\sqrt3}& \frac{1}{\sqrt3}&\frac{1}{\sqrt3} \\[0.7em]
0 & \sin\alpha & \cos\alpha
\end{pmatrix} \label{eq:PMNSForm}
 }
 has led to an $A_4$ symmetry, as shown analytically in  \cite{FK19}. The key points of Ref. \cite{FK19} are 
\begin{itemize}
\item[{\bf 1}] We choose the bases such that $\Qem=-\frac{1}{3}$ quarks and $\Qem=-1$ leptons are mass eigenstates. 
\item[{\bf 2}] $\ell_L$ in $\five$ of  Eq. (\ref{eq:DefF5}) is a triplet under the tetrahedral group $A_4$.
\item[{\bf 3}] All quark states in  Eq. (\ref{eq:DefF5}) are singlets  under $A_4$.
\item[{\bf 4}] The Higgs doublet(s) is a singlet under the permutation symmetry group $A_4$.
\end{itemize}

\bigskip

\noindent 
There are four representations in $A_4$: $\three, \one, \one',$ and $\one''$.
Let us remark first that Item {\bf 1} evades the problem encountered in the Georgi-Jarlskog relation. We choose the needed mass values in the definition of the  $\Qem=-\frac{1}{3}$ quark masses.  Item {\bf 4} requires that the Higgs quintet $\fiveb^H_{+2}$ is a tetrahedral group singlet. Then,  Items {\bf 2} and {\bf 3} dictate to assign $u^c_L$ in the triplet representation {\bf 3} of $A_4$ since both $\ell_L$  and $u^c_L$ belongs to the same representation $\five_{+3}$. 

\bigskip

\noindent 
The tensor product of two {\bf 3}'s of $A_4$ is
\dis{
{\three}\otimes {\three}= 2\cdot\three\oplus \one\oplus \one'\oplus \one''.
\label{eq:A4Tensor}
}
We use the representations where three $\Qem=-\frac13$ quarks of  each chirality form a representation {\bf 3} of $A_4$, so do charged leptons. Then, the tensor product Eq. (\ref{eq:A4Tensor}) allows three parameters, viz. three singlets, for three  $\Qem=-\frac13$ quark masses and choosing the diagonal basis for  $\Qem=-\frac13$ quarks is guaranteed from $A_4$. The same applies to charged leptons also.

\bigskip

\noindent
 Note that the charged currents(CCs) in the SM are given by 
\dis{
 \frac{g}{\sqrt2}\left( \bar{u}^{(\rm 0)}_{L} \gamma^\mu d^{(\rm mass)}_{L}  +   \bar{\nu}^{(\rm 0)}_{L} \gamma^\mu e^{(\rm mass)}_{L}\right) W^+_\mu+{\rm h.c.}
}
where
\dis{
{u}^{(\rm 0)}_{L} =\begin{pmatrix} u^{(0)}\\[0.3em] c^{(0)}\\[0.3em] t^{(0)}\end{pmatrix}_L,~  {\nu}^{(\rm 0)}_{L} =\begin{pmatrix} \nu_e^{(0)}\\[0.3em] \nu_\mu^{(0)}\\[0.3em] \nu_\tau^{(0)}\end{pmatrix}_L.\label{eq:0andi}
}
With the anti-SU(5) representations of (\ref{eq:DefF5}), these CC's are included in  
\dis{
 g\left( \bar{\tenb}_{-1} \gamma^\mu T_{\bar{10}}^{-}\tenb_{-1} + \bar{\five}_{+3} \gamma^\mu  T_5^-\five_{+3}\right) W^+_\mu+{\rm h.c.}\label{eq:CCgut}
}
where
\dis{
T_5^-=\begin{pmatrix} 0&0&0&0&0\\ 0&0&0&0&0\\ 0&0&0&0&0\\ 0&0&0&0&\frac{1}{\sqrt2}\\ 0&0&0&0&0
\end{pmatrix}
}
and $T_{\bar{10}}^{-}$ changes $d^\alpha$ to $u^\alpha$. Three families are
\dis{
\overline{\bf T}=\begin{pmatrix}\tenb_{-1}^d,\, \tenb_{-1}^s,\,\tenb_{-1}^b\end{pmatrix}, ~~{\bf F}=\begin{pmatrix}\five_{+3}^e,\, \five_{+3}^\mu,\,\five_{+3}^\tau \end{pmatrix}
}
where $d,s,b$ and $e,\mu,\tau$ are family indices.
In terms of mass eigenstates  $\Qem=+\frac23$ quarks ($u,c,t$) and  neutrinos  ($\nu_1,\nu_2,\nu_3$),  the weak eigenstates of (\ref{eq:0andi})  are related by L-sector unitary matrices $U$ and  R-sector unitary matrices  ${\cal U}$ by
\dis{
\begin{pmatrix} u^{(0)}\\[0.3em] c^{(0)}\\[0.3em] t^{(0)}\end{pmatrix}_L= U^{(u)\,\dagger}\begin{pmatrix} u \\[0.3em] c \\[0.3em] t \end{pmatrix}_L,~~  \begin{pmatrix} \nu_e^{(0)}\\[0.3em] \nu_\mu^{(0)}\\[0.3em] \nu_\tau^{(0)}\end{pmatrix}_L= U^{(\nu)\,\dagger}\begin{pmatrix} \nu_1\\[0.3em] \nu_2\\[0.3em] \nu_3\end{pmatrix}_L,~~  \begin{pmatrix} N_e^{(0)}\\[0.3em] N_\mu^{(0)}\\[0.3em] N_\tau^{(0)}\end{pmatrix}_R= {\cal U}^{(\nu)\,\dagger}\begin{pmatrix} \nu_1\\[0.3em] \nu_2\\[0.3em] \nu_3\end{pmatrix}_R.
}

\bigskip
\noindent
Now,  Eq. (\ref{eq:CCgut}) reads for three families as
\dis{
 g\left(\bar{\overline{\bf T}} \gamma^\mu T_{\bar{10}}^{-}\overline{\bf T} + \bar{{\bf F}} \gamma^\mu  T_5^-{\bf F} \right) W^+_\mu+{\rm h.c.}\label{eq:CCgutF}
}
The CKM and PMNS matrices are given by
\dis{
V^{\rm (CKM)}= U^{(u)} U^{(d)\,\dagger}= U^{(u)} ,~V^{\rm (PMNS)}=U^{(\nu)} U^{(e)\,\dagger}= U^{(\nu)} .\label{eq:VandU}
}
The definitions of $ U^{(u)}$ and $U^{(\nu)}$ in Eq. (\ref{eq:VandU}) have the required number of parameters. In  $ U^{(u)}$, there are just two phases of L-handed $ u^{(0)}$ quarks for constraints because the baryon number phase cannot be used as a constraint. Also, three  $ u^{(0)}$ masses provide three constraints. Thus, out of 9 parameters in a $3\times 3$ unitary matrix, the number of undtermined parameters are 4: 3 real angles and 1 phase. In  $ U^{(\nu)}$, we do not have any phase constraint because Majorana neutrinos are real. So, we have nine parameters minus three mass parameters, leading to 3 real angles, 1 Dirac phase and 2 Majorana phases.

\bigskip
\noindent
 Let us consider the leptonic part first, which is included in the 2nd term in Eq. (\ref{eq:CCgutF}). Since neutrinos belong to the triplet representation of $A_4$, ${\bf F}$ transforms as $\three$ under $A_4$. The $\Qem=-1$ leptons being chosen as mass eigenstates, there remains to choose $\nu^{(0)}$. Thus, the $A_4$ symmetric property of $\bar{\bf F}\otimes {\bf F}$ is $\one\oplus \one'\oplus  \one''\oplus 2\cdot\three $, from which  we choose $\one\oplus \one'\oplus  \one''$ for Eq. (\ref{eq:CCgutF}) to be $A_4$ symmetric. Thus, {\bf F} can be chosen as
 \dis{
 {\bf F}^{(0)} \ni a \nu_e^{(0)},~b \nu_\mu^{(0)},~c \nu_\tau^{(0)},
 }
 which are matched with charged leptons $e,\mu,$ and $\tau$.

\bigskip
\noindent
In the quark sector,  quarks are treated as singlets $\one, \one'$ and $\one''$. So, the first term of Eq. (\ref{eq:CCgutF})
is $A_4$ symmetric.  With these CC couplings, the question to discuss next is how the quark and lepton Yukawa couplings are given.
  
\section{Yukawa couplings}\label{sec:Yukawa} 
 
\begin{figure}[!h]
\hskip 0.01cm \includegraphics[width=0.5\textwidth]{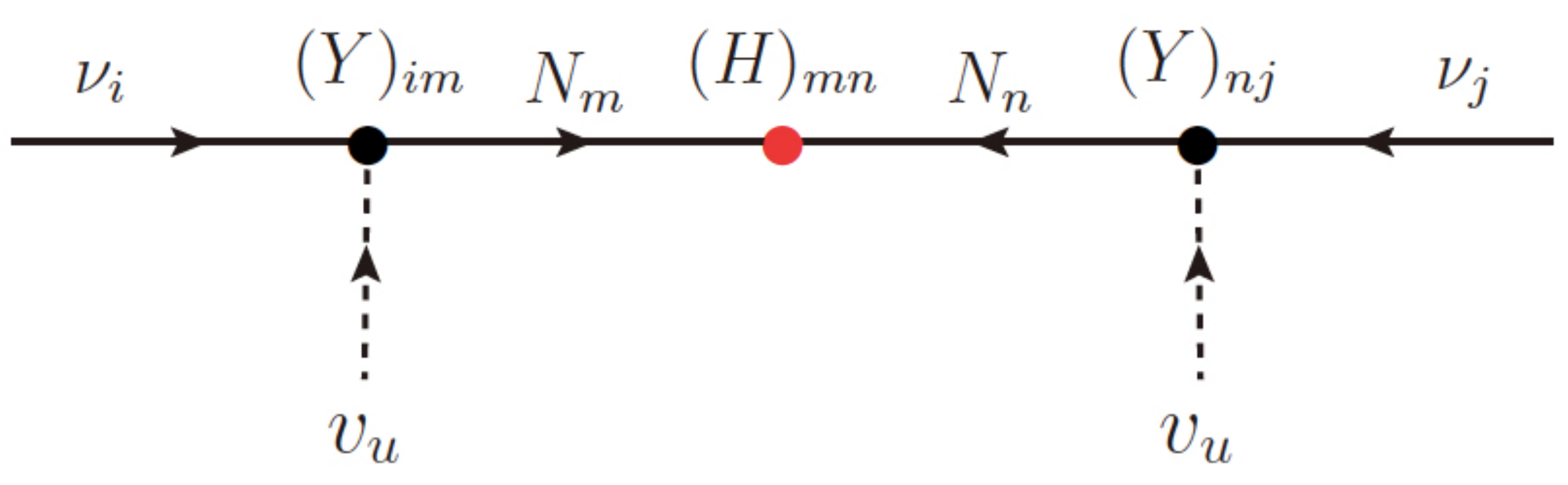}
\caption{The Type 1 see-saw diagram. }\label{fig:Seesaw}
\end{figure}

\noindent
To realise $A_4$ symmetry, we assign the Yukawa couplings such that the flavor indices of $i$ respect the $A_4$ symmetry  requirements. Since the $A_4$ symmetry was suggested from the PMNS matrix, let us first discuss the L violating neutrino masses. Since ${\bf F}$ is complex, it can have a global U(1) phase which is not violated by Eq. (\ref{eq:CCgutF}). The charged lepton in {\bf F} obtains mass by the Yukawa coupling to $\oneb_{-5}=e^c$ of Eq. (\ref{eq:DefF5}), $\oneb_{-5} C^{-1}{\bf F}_{+3}\fiveb^H_{+2}$. Since $e^c$, \ie $\oneb_{-5}$, carries lepton number L=--1, ${\bf F}_{+3}$ carries L=+1. But  ${\bf F}_{+3}$   also  contains $u^c$ which is known to carry baryon number B=--1. For consistency, we require no global anomaly. So,    ${\bf F}_{+3}$  should carry a vanishing global charge which can be (B--L).   ${\bf F}_{+3}$  couples to  $\bar{\bf T}_{-1}$ by $\bar{\bf T}_{-1}C^{-1} {\bf F}_{+3}\,\five^H_{-2}$. Since $\five^H_{-2}$ is interpreted carrying no B and L charges, $\bar{\bf T}_{-1}$ carries B=+1 or  L=--1. In particular $N_L^c$ carries L=--1. Namely, $N_R$ carries L=+1. 
The L violating source at the super-renormalizable level is given by $({m_N}/{2})N_R^2$. What is the $A_4$ representation of $\bar{\bf T}_{-1}$? To write  $\frac{m_N}{2} (N_R)^2$,  $\bar{\bf T}_{-1}$ transforms as a singlet(s) or {\bf 3} of $A_4$. These L violating heavy neutrino masses are contained in
\dis{
(\bar{\bf T}_{-1})_i (H)_{ij}(\bar{\bf T}_{-1})_j,
}
where $(H)$ is the heavy neutrino mass matrix. Since we do not introduce any triplet in the Higgs or fermion sectors, our neutrino mass matrix will be a Type 1 see-saw. The Dirac neutrino mass is given by
\dis{
{\bf F}_{+3\,i}\,Y_{ij}\,\bar{\bf T}_{-1 ,j} \five^H_{-2}
\label{eq:YukawaF5}
}
where $(Y)$ is the Yukawa coupling matrix. In Fig. \ref{fig:Seesaw}, we show the tree diagram for the Type 1 see-saw mechanism. In Fig. \ref{fig:Seesaw}, the chiralities of $\nu$ and $N$ are L and R, respectively. This diagram depends on the $A_4$ property of $N$. 
With these diagrams, we obtain the effective Weinberg operators,
\dis{
 h_{ij}\frac{v_u^2}{m_N} \, \tilde{\nu}^{(0)}  _{iL}C^{-1}\nu^{(0)}  _{jL}.\label{eq:WeinNu}
}
 
 \bigskip
 
\begin{figure}[!h]
\hskip 0.01cm \includegraphics[width=0.5\textwidth]{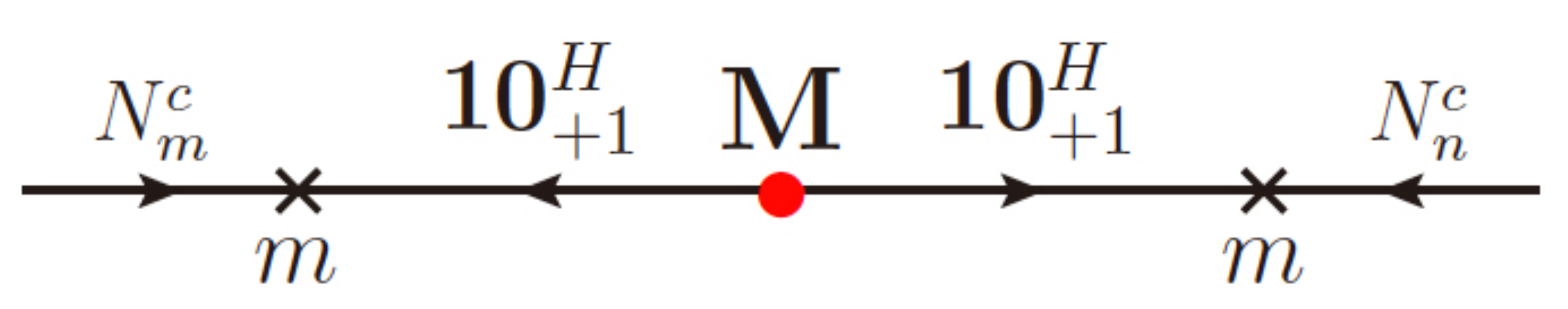}
\caption{The  diagram for heavy neutrino masses. The fermionic partner of the GUT Higgs $\ten^H_{+1}$, \ie $B_{45}$ of Eq. (\ref{eq:GUTHiggs}), is called ``heavy heavy neutrino''.}\label{fig:HeavyNMass}
\end{figure}

 \noindent
 We noted above that $\bar{\bf T}_{-1}$ of Eq. (\ref{eq:DefF5}) transforms as {\bf 3} under $A_4$, so does $N^c_L$ in  $\bar{\bf T}_{-1}$. The Yukawa coupling in  
 $(H)$ of Fig. \ref{fig:Seesaw} is a constant because all three $N_R$'s belong to {\bf 3} of $A_4$. But, we allow the difference among masses of three $N$. Thus, $(H^{-1})$ is just the inverse of the mass matrix of $N$. But, the mass term of $N$ cannot arise at the renomalizable level. It occurs only through the dimension-5 term, from fields in Eqs. (\ref{eq:GUTHiggs}) and (\ref{eq:DefF5}),
 \dis{
 \sim \bar{\bf T}_{-1}({\rm fermion})\bar{\bf T}_{-1}({\rm fermion})\ten^H_{+1}({\rm boson})\ten^H_{+1}({\rm boson})\label{eq:Wein5}
 }
where the VEV $\langle\ten^H_{+1}({\rm boson})\rangle$is needed to break the anti-SU(5) to the SM gauge group.
The gauge invariant super-renormalizable mass term breaking lepton number L is
\dis{
 \bar{\bf T}_{-1}({\rm fermion}) {\bf m}\ten^H_{+1}({\rm fermion})\label{eq:mTerm}
}
where ${\bf m}$ is a constant (or matrix). This dictates that $\ten^H_{+1}({\rm fermion})$ of Eq. (\ref{eq:mTerm}) transforms as  {\bf 3} of $A_4$. In Fig. \ref{fig:HeavyNMass}, we draw a schematic Feynman diagram generating the heavy neutrino masses from the anti-SU(5)$\times A_4$ symmetry.\footnote{Identifying  $\ten^H_{+1}$'s of Eqs. (\ref{eq:Wein5}) and (\ref{eq:mTerm}), we may be led to introduce supersymmetry.}
The mass matrix ${\bf M}$ transforms, under $A_4$,  as $\three\otimes \three$,   $\three\otimes \one$, $\one\otimes \three$,  or $\one$'s, where the left factor combines with $N_m^c$ and  the right factor combines with $N_n^c$. For each case, we study the L-violating neutrino masses.

 \bigskip
 \noindent
 Before discussing  each  neutrino mass matrix, we present  the $\Qem=+\frac23$ quark masses from the anti-SU(5) coupling which depends only on the coupling given in Eq. (\ref{eq:YukawaF5}). The $\Qem=+\frac23$ quark Yukawa couplings  are determined from the $A_4$ tensor product ${\three}\otimes{\three}={\one}\oplus{\one'}\oplus{\one''}\oplus {2\cdot\three}$. There are three independent singlets, which are three independent Yukawa couplings. $Y_{ij}$  in Eq. (\ref{eq:YukawaF5}) are matrix elements. There is only on class for matrices which have Det$=1$ and Tr$=-1$ for entries with $\pm 1$,  
\dis{
\begin{pmatrix} -1&0&0\\ 0&-1&0\\ 0& 0& 1 \end{pmatrix}.\label{eq:matrixform}
}
For example, the matrix
\dis{
\begin{pmatrix} 0&-1&0\\ 1&0&0\\ 0& 0& -1 \end{pmatrix}
}
satisfies the required conditions but changing the indices $1\leftrightarrow 2$ gives the form Eq. (\ref{eq:matrixform}). Similarly, all the other cases can be reduced to the form (\ref{eq:matrixform}). For  Eq. (\ref{eq:matrixform}), the Yukawa couplings are defined as $Y_{11}=h_1, Y_{22}=h_2, Y_{33}=h_3$, and all the rest are zeros.

 \bigskip
 \noindent
Now let us proceed to discuss each class of {\bf M} on neutrino masses.

\subsection{ ${\bf M}\sim \three\otimes \three$ }

\bigskip

\noindent
 In this case, three values are the same for the left and right factors. In the matrix form,
 \dis{
 {\bf M}=\begin{pmatrix} M& M& M\\ M& M & M\\ M & M & M
 \end{pmatrix}
 }
 which has eigenvalues of $3M, ~0,$ and 0. The above is a democratic form suggested in Refs. \cite{Fritzsch96,Fritzsch04}. The heavy neutrino mass components are
 \dis{
 {\bf M}_{mn}=\frac{m^2}{M}.
 }
 In this case $Y_{ij}$ of Eq. (\ref{eq:YukawaF5}) is $Y_{ij}=h\delta_{ij}$.
Then, the SM neutrinos obtain masses through Fig. \ref{fig:Seesaw},
 \dis{
 m_{ij}=h^2\frac{Mv_u^2}{m^2}.\label{eq:DemoForm}
 }
The above universal mass matrix is diagonalised by
\dis{
 \begin{pmatrix} \frac{\sqrt3-1}{2\sqrt3} &  \frac{-\sqrt3-1}{2\sqrt3}&\frac1{\sqrt3}
\\[0.5em]
 \frac{-\sqrt3-1}{2\sqrt3}& \frac{\sqrt3-1}{2\sqrt3}&\frac1{\sqrt3}
\\[0.5em]
\frac1{\sqrt3}&\frac1{\sqrt3}&\frac1{\sqrt3}\label{eq:PMNSTry}
\end{pmatrix}
} 
which is tri-maximal.

\subsection{${\bf M}\sim\three\otimes \one$ }

\bigskip

\noindent
The left factors give the same value and and the right factors give three different values. 
  \dis{
 {\bf M}=\begin{pmatrix} M_1& M_2& M_3\\ M_1& M_2 & M_3\\ M_1 & M_2 & M_3
 \end{pmatrix}
 }
 which has eigenvalues of $M_1+M_2+M_3, ~0,$ and 0. All the heavy neutrinos have the same mass,
  \dis{
 {\bf M}_{mn}=\frac{m^2}{M_n}.
 }
  In this case $Y_{ij}$ of Eq. (\ref{eq:YukawaF5}) is $Y_{ij}=h\delta_{ij}$ at the LHS vertex and $h_j$ at the RHS vertex. Thus, the SM neutrinos obtain masses through Fig. \ref{fig:Seesaw} as
 \dis{
 m_{ij}=hh_j \frac{M_{j}v_u^2}{m^2},
 }
 which is proportional to
  \dis{
m\propto \begin{pmatrix} h_1& h_2& h_3\\  h_1& h_2& h_3\\  h_1& h_2& h_3
 \end{pmatrix}\label{eq:mForm31}
 }
 whose eigenvalues are 0, 0, and $h_1+h_2+h_3$. Three  column vectors of $m^T$ with eigenvalues 0, 0, and $h_1+h_2+h_3$ are
  \dis{
 \psi_1\sim \begin{pmatrix} h_2\\[0.3em]  -h_1 \\[0.3em] 0
 \end{pmatrix}, 
 \psi_2\sim\begin{pmatrix}h_1\\[0.3em]   h_2\\[0.5em]  \frac{-h_1^2-h_2^2}{h_3} 
 \end{pmatrix},
  \psi_3\sim\begin{pmatrix}  h_1\\[0.3em]   h_2\\[0.3em]   h_3
 \end{pmatrix}.
 }
 Note that Eq. (\ref{eq:mForm31}) has a freedom to choose the scale. We fix such that the unitarity matrix results. The unitarity matrix diagonalizing $m^T$ is
 \dis{
U=\frac{1}{\sqrt{h^2(1+h^2)}} \begin{pmatrix} h_2\sqrt{1+h^2}&  h_1 &  h_1h\\[0.3em]  
 -h_1\sqrt{1+h^2}& h_2&   h_2h \\[0.3em]   
0& -h^2  & h
 \end{pmatrix},
 }
where we choose  
 \dis{ &h_3=1,\\
& h^2\equiv h_1^2+h_2^2.
 }
 Then, the diagonalized states and matrix are expressed in terms of the original ones as
 \dis{
& \psi^{(\rm diag)}=U\psi_0,\\
 &m^{(\rm diag)}=U m\, U^\dagger.
 }
 
\subsection{${\bf M}\sim\one\otimes \three$ }

\bigskip

\noindent
The left factors give three different value and and the right factors give the  same values. 
  \dis{
 {\bf M}=\begin{pmatrix} M_1& M_1& M_1\\ M_2& M_2 & M_2\\ M_3 & M_3 & M_3
 \end{pmatrix}
 }
 which has eigenvalues of $3M, 0,$ and 0. All the heavy neutrinos have the same mass,
\dis{
 {\bf M}_{mn}=\frac{m^2}{M_m}.
 }

\bigskip

\noindent 
 In this case $Y_{ij}$ of Eq. (\ref{eq:YukawaF5}) is $Y_{ij}=h_i$ at the LHS vertex and $h\delta_{ij}$ at the RHS vertex. Thus, the SM neutrinos obtain masses through Fig. \ref{fig:Seesaw} as
 \dis{
 m_{ij}=hh_i \frac{M_{i}v_u^2}{m^2}.
 }
 As in Case {\bf B}, we obtain the following
  \dis{
U=\frac{1}{\sqrt{h^2(1+h^2)}} \begin{pmatrix} h_2\sqrt{1+h^2}&  -h_1\sqrt{1+h^2}&  0\\[0.3em]  
 h_1 & h_2&   -h^2 \\[0.3em]   
h_1h& h_2h  & h
 \end{pmatrix},
 } 
 where
 \dis{
& \psi^{(\rm diag)}=U\psi_0,\\
 &m^{(\rm diag)}=U m\, U^\dagger.
 }

\subsection{${\bf M}\sim\one$'s }
\noindent
  In this case, both the left and right factors give three different values. 
  \dis{
 {\bf M}=\begin{pmatrix} M_{11}& M_{12}& M_{13}\\ M_{21}& M_{22}& M_{23}\\ M_{31} & M_{32} & M_{33}
 \end{pmatrix}
 }
 which in general gives three different nonzero eigenvalues. All the heavy neutrinos have the same mass,
 \dis{
 {\bf M}_{ij}=\frac{m^2}{M_{ij}}.
 }
  In this case $Y_{ij}$ of Eq. (\ref{eq:YukawaF5}) is $Y_{ij}=h_i$ at the LHS vertex and $h_j$ at the RHS vertex. Thus, the SM neutrinos obtain masses through Fig. \ref{fig:Seesaw} as
 \dis{
 m_{ij}=h_ih_j \frac{M_{ij}v_u^2}{m^2},
 }
 which is general enough to obtain any unitarity matrix $U$.

\subsection{Allowed matrices for {\bf M} from effective neutrino masses}

\noindent  
In the above subsections, the  heavy heavy neutrno mass matrix {\bf M} of Fig. \ref{fig:HeavyNMass}, leading to the heavy neutrino masses of $N$ in Eq. (\ref{eq:DefF5},) were given. On the other hand, the effective neutrino mass operator of Weinberg \cite{Weinberg79},
\dis{
\sim \ell_i^T C^{-1}\ell_j\label{eq:Weinberg}
}
is symmetric on the exchange $i \leftrightarrow j$. But, Cases {\bf B} and  {\bf C} allow asymmetric neutrino masses. Therefore, the  heavy heavy neutrno mass matrix {\bf M} can take only Cases {\bf A} and  {\bf D}. Since Case {\bf D} is not very much predictive at this stage, we present the $A_4$ from the anti-SU(5) prediction given in Eq. (\ref{eq:PMNSTry}),
\dis{
 \begin{pmatrix}   \frac{-\sqrt3-1}{2\sqrt3}&\frac1{\sqrt3}&\frac{\sqrt3-1}{2\sqrt3} 
\\[0.5em]
 \frac1{\sqrt3}&\frac1{\sqrt3}&\frac1{\sqrt3} 
\\[0.5em]
 \frac{\sqrt3-1}{2\sqrt3}&\frac1{\sqrt3}&\frac{-\sqrt3-1}{2\sqrt3}
\end{pmatrix}\label{eq:PMNSDemo}
} 
where $|(U^\dagger)_{13}|=\frac{\sqrt3-1}{2\sqrt3}\simeq 0.211$.
The best fit \cite{PDG18PMNS} gives  0.147 and $3\sigma$ range is $0.138-0.156$. Therefore, Case {\bf A} is ruled out. Only, Case {\bf D}, which is general enough, is a viable mass pattern of the heavy heavy neutrinos.
\subsection{The CKM matrix}

\noindent
For Case {\bf D}, let us consider the CKM matrix. In Ref. \cite{FK19}, we argued that the CKM matrix is close to the identity because of the huge ratio of $m_t/m_c$.  So, the mass matrix is of the form, 
\dis{
\sim\begin{pmatrix}
a\varepsilon^2  &b\varepsilon^{\frac{3}{2}}  & c\varepsilon  \\
d\varepsilon^{\frac{3}{2}}  & e\varepsilon  & f\varepsilon^{\frac{1}{2}}  \\
 g\varepsilon &h\varepsilon^{\frac{1}{2}} & 1
\end{pmatrix}
 }
where $\varepsilon$ is O$(\frac{m_c}{m_t})\approx 0.007$. The determinent of the above matrix is  $D=(ae-bd +bfg+cdh-afh-ceg)\varepsilon^3$. Choose $e\simeq 1$ such that trace is almost $m_t+m_c$. $D\simeq (a-bd-cg +bfg+cdh-afh)\varepsilon^3$, and hence $m_u\simeq m_c(m_c/m_t)^2(a-bd-cg +bfg+cdh-afh)\simeq 2.5\meV$ leading to $(a-bd-cg +bfg+cdh-afh)\simeq 43$. Since we follow Case {\bf D}, all these coefficients $a\sim g$ are arbitrary. Let us take a real symmetric matrix,  choosing simple numbers   just for an illustration,
 \dis{
  a=-32.1615, e=1,b=d=c=g=(43)^{1/3}, f=h=1.42, \label{eq:abcset}
 }
where $a$ is chosen to satisfy $(a-bd-cg +bfg+cdh-afh)\simeq 43$.
In this case, the mass matrix is \cite{KimKim20},
 \dis{
 M\sim \left(
\begin{array}{ccc}
 -0.00157596, & 0.00205181, & 0.0245238 \\
 0.00205181, & 0.007, & 0.118806 \\
 0.0245238, & 0.118806, & 1 \\
\end{array}
\right).\label{eq:NumInput}
 }
 Then,  eigenvalues of $M$ are  
 \dis{
 -0.00203125,   -0.00715645,   1.01461,
 }    
 where the first term can be corrected more by higher dimensional operators. Here, $m_c/m_t\simeq 0.007$, and the diagonalizing matrix, $UMU^\dagger=$(diagonal),  is
\dis{
 V^{\rm(CKM)}=U^{(u)}= 
\left(
\begin{array}{ccc}
 0.985959 & -0.166933 & -0.00433802 \\
 -0.165227 & -0.978988 & 0.119506 \\
 0.0241964 & 0.117111 & 0.992824 \\
\end{array}
\right)
 } 
 which gives the Cabibbo angle $|\theta_C|\simeq  9.61^{\rm o}$, roughly 3.4$^{\rm o}$ smaller than the needed one. Note however that  we neglected the CP  phase $\delta$ and other higher dimensional contributions. Most importantly, it is for a specific set of parameters in Eq. (\ref{eq:abcset}). In general, the mass matrix is complex which can be diagonalized by bi-unitary matrices, by  $U$ and ${\cal U}$.   In sum, we tried to show there can be a reasonable set of parameters  fitting all the flavor data for Case {\bf D}.
  
\color{black}

\section{String compactification}\label{sec:Z12Rp} 
 
\bigskip

\noindent
To discuss flavor symmetry from string compactification, one needs a compactification model where details of the SM field assignment is presented. In doing so, the key SM phenomenologies are automatically included, \ie it is not ruled out from any well established data. Here, we show a realisation of $A_4$ symmetry based on an  anti-SU(5) GUT \cite{KimRp19} possessing the $\Z_{4R}$ discrete parity  which is obtained from the $\Z_{12-I}$ compactification of  the $\EE8$ heterotic string \cite{Gross85}.  Anyway, for a detail study of flavor physics, one has to specify every aspect of the flavors for which we do not find any reference except Ref. \cite{KimRp19}. So, we show an example of $A_4$ symmetry based on an anti-SUY(5) GUT of \cite{KimRp19} based on the model \cite{Huh09}.   Here, we just cite the needed information from Refs.  \cite{KimRp19,Huh09}. In string compactification, the needed Yukawa couplings arise by satisfying all the selection criteria. 
The anti-SUY(5) GUT of \cite{KimRp19} does not allow any SM Yukawa couplings at the renormalizable level. But, at the level of dimension-5 there appear the SM Yukawa couplings which are proportional to the VEVs of $\langle \ten^H_{+1}\rangle= \langle \tenb^H_{-1}\rangle$. Since these VEVs are near the string scale, we obtain top quark mass at the order the electroweak scale. Since we are not attempting to discuss details of models in string compactification, we only pay attention to the multiplicities of the needed chiral fields.
 
\bigskip
\noindent
First consider $\ten_{+1}^H$ and  $\tenb_{-1}^H$ needed for breaking anti-SU(5). In the $T_3$ twisted sector, chiral fields are contructed in Eqs. (23) and (24) of Ref.   \cite{KimRp19},
\dis{
 \begin{array}{cccc} 
 {s}& {\rm Multiplicity} &P\cdot V\\[0.3em]
 (\oplus|-+-): &  2,&\frac{+1}{4} (\Sigma_1^*)\\[0.3em] 
 (\ominus|---) &1,&\frac{-1}{4}(\Sigma_2) \\[0.3em] 
  \end{array}  \label{eq:HeavyHA}
}
 \dis{
 \begin{array}{ccccccc} 
 {s} &{\rm Multiplicity}  &P\cdot V\\[0.3em]
 (\oplus|-+-): &  1, &\frac{+1}{4} (\Sigma_1^*)\\[0.3em] 
 (\ominus|---): &    2, &\frac{-1}{4}(\Sigma_2) \\[0.3em] 
  \end{array}  \label{eq:HeavyHB}
}
where $\oplus$ and $\ominus$ denote L-handed and R-handed chiral fields respectively.  So, here we consider only the number of chiral fields at the same fixed points. We cited only chirality and multiplicity.   $\Theta_i$ in Table I and $P\cdot V$ in Eqs. (\ref{eq:HeavyHA}) and (\ref{eq:HeavyHB})  are used to calculate the multiplicity. From Eqs. (\ref{eq:HeavyHA}) and (\ref{eq:HeavyHB}), note that there appear three L-handed fields $\Sigma_2$, and  three R-handed fields $\Sigma_1^*$. These chiral fields at the same fixed points are not distinguished. Thus, the L-handed fields $\Sigma_2$ has the representation ${\bf 3}$ of $A_4$, so do  the R-handed fields $\Sigma_1^*$.   $\Sigma_2$ is the one for $\ten_{+1}^H$ of Eq. (\ref{eq:mTerm}). But  $\bar{\bf T}_{-1} $ of Eq. (\ref{eq:mTerm}) belongs to the matter fields in Table I of \cite{KimRp19}. Two matter $\tenb_{-1} $'s appear in $T_4^0$, viz. Table I.  But it is better to check all $\tenb_{-1} $'s before removing vectorlike representations, for which we go back to Ref. \cite{Huh09}. In fact, there was no vectorlike representations of  $\tenb_{-1}\oplus \ten_{+1} $'s removed in Ref.  \cite{Huh09}. So, from our string model, $\tenb_{-1}$ is a doublet ${\bf 2}$ of permutation symmetry $S_3$. We do not realize the coupling of Eq.   (\ref{eq:mTerm}).

\bigskip

\begin{table}[t!]
\begin{center}
\begin{tabular}{@{}lc|cc|c|cccccc|ccc@{}} \toprule
 &State($P+kV_0$)&$~\Theta_i~$ &${\bf R}_X$(Sect.) \\[0.1em] \colrule
 $\xi_3$  & $(\underline{+++--};--+)(0^8)'$&$0$ &$\tenb_{-1}(U_3)$  \\
$\bar{\eta}_3$  & $(\underline{+----};+--)(0^8)'$&$0$ &$\five_{+3}(U_3)$  \\
$\tau^c$  & $({+++++};-+-)(0^8)'$& $0$ &$\one_{-5}(U_3)$   \\
$\xi_2$  & $(\underline{+++--};-\frac{1}{6},-\frac{1}{6},-\frac{1}{6})(0^8)'$& $\frac{+1}{4}$ &$\tenb_{-1}(T_4^0)$ \\
$\bar{\eta}_2$  & $(\underline{+----};-\frac{1}{6},-\frac{1}{6},-\frac{1}{6})(0^8)'$&$\frac{+1}{4}$ &$\five_{+3}(T_4^0)$ \\
$\mu^c$  & $({+++++};-\frac{1}{6},-\frac{1}{6},-\frac{1}{6})(0^8)'$&$\frac{+1}{4}$ & $\one_{-5}(T_4^0)$ \\
$\xi_1$  & $(\underline{+++--};-\frac{1}{6},-\frac{1}{6},-\frac{1}{6})(0^8)'$&$\frac{+1}{4}$ &$\tenb_{-1}(T_4^0)$ \\
$\bar{\eta}_1$  & $(\underline{+----};-\frac{1}{6},-\frac{1}{6},-\frac{1}{6})(0^8)'$&$\frac{+1}{4}$ &$\five_{+3}(T_4^0)$ \\
$e^c$  & $({+++++};-\frac{1}{6},-\frac{1}{6},-\frac{1}{6})(0^8)'$&$\frac{+1}{4}$ & $\one_{-5}(T_4^0)$ \\[0.2em]
\hline
 $H_{uL}$  & $(\underline{+1\,0\,0\,0\,0};\,0\,0\,0)(0^5;\frac{-1}{2}\,\frac{+1}{2}\,0)'$&$\frac{+1}{3}$ &$2\cdot \five_{-2}(T_6)$  \\
 $H_{dL}$  & $(\underline{-1\,0\,0\,0\,0};\,0\,0\,0)(0^5;\frac{+1}{2}\,\frac{-1}{2}\,0 )'$&$\frac{+1}{3}$ &$ 2\cdot \fiveb_{+2}(T_6)$   \\ [0.2em]
 \botrule
 \end{tabular} \label{tab:SMqn}
\end{center}
\caption{Phases $\Theta_i$ of matter fields in the SM.  $U$ and $T$ are twisted sectors. In $T_4^0$, there are two $\tenb_{-1} $'s.}
\end{table}

 \begin{table}[!h] 
\caption{Branching of $S_4$ representations $\one, \one',\two, \three$ and $\three'$ into the $A_4$ and $S_3$ representations \cite{KobaBk12}.}
\begin{tabular}{c|c|c}
\hline
 &\\[-1.3em]  
 $S_4$~& $A_4$ & $S_3$\\ \hline
   $\one$~ & $\one$&$\one$  \\[0.3em]
  $\one'$ &$\one$&$\one'$  \\[0.3em]
  $\two$~ &~~$\one'\oplus \one''$~~&$\two$  \\[0.3em]
  $\three$~ &$\three$&$\one\oplus \two$  \\[0.3em]
  ~$\three'$~~&$\three$&~$\one'\oplus \two$~\\[0.3em]
  \hline
  \end{tabular}\label{tab:S4A4} 
\end{table}

\bigskip

\noindent
From Table \ref{tab:S4A4}, we note that the doublet representation {\bf 2} of the permutation group $S_3$ can be obtained from {\bf 3} of $S_4$. Also,  {\bf 3} of $A_4$ is  from {\bf 3} of $S_4$.

\bigskip

\noindent
In Eq.   (\ref{eq:mTerm}),  $\bar{\bf T}_{-1}$ transforms as    {\bf 2} under the permutation group $S_3$ and $\ten^H_{+1}$  transforms as   {\bf 3}  of $A_4$. Note that {\bf 2} and {\bf 3} of $S_4$ produce  {\bf 2} of $S_3$. Out of two {\bf 1}'s of $S_4$, we restrict to only  {\bf 1}.    Let us consider the relevant tensor products of $S_4$,
\dis{
\textrm{Tensor products in }S_4\left\{\begin{array}{l}
\three\times \three=\one\oplus \two\oplus \three\oplus \three'\\
\three\times \three'=\one'\oplus \two\oplus \three\oplus \three'\\
\three'\times \three'=\one\oplus \two\oplus \three\oplus \three'\\
 \two\otimes \three= \three\oplus \three' 
\end{array}\right.
\label{eq:S4products}
}
where the last line does not produce a singlet. The other three lines produce $S_4$ singlets and we consider the first line, $\three\otimes\three$. The other cases can be equivalent to this by redefining the origin of {\bf 3} of $A_4$. Then, $\bar{\bf T}_{-1}$ and $\ten^H_{+1}$  can be traced back  to {\bf 3} of $S_4$.
\dis{
R(S_3) \otimes R(A_4): &(\one, \two)\otimes  \three\\
 \to &(\one\otimes\three) \oplus (\two\otimes\three)\\
 \to &\one\otimes\three\oplus \one'\otimes\three\oplus \one''\otimes\three 
}
where the first line is the fourth line of Table \ref{tab:S4A4}, written as $S_3$ and $A_4$ subgroups. In the 2nd line, $\one\otimes\three$ is the $S_4$ product and can be interpreted as the $A_4$ triplet.  In the 2nd line, $\two\otimes\three$ is $S_4$ product from the 3rd and 4th lines of  Table \ref{tab:S4A4}. In terms of $A_4$, it produces $\one'\otimes\three\oplus \one''\otimes\three$, \ie two independent $\three$'s. In total, there are three independent $\three$'s.   Therefore, ${\bf M}$ of Fig. \ref{fig:HeavyNMass} is 
 \dis{
 {\bf M}=\begin{pmatrix} M_{11}& M_{12}& M_{13}\\ M_{21}& M_{21} & M_{23}\\ M_{31} & M_{32} & M_{33}
 \end{pmatrix}
 }
 which is Case {\bf D} of Sec. \ref{sec:Yukawa}, which is allowed from the neutrino mass data.

\bigskip

\noindent
So far we paid attention to the heavy neutrino mass in $\bar{\bf T}_{-1}$. Now, let us check how this representation containing a quark doublet predicts on the Yukawa couplings through Eq. (\ref{eq:YukawaF5}) with a R-handed $\Qem=\frac23$ quark in ${\bf F}_{+3}$. Both
 $\bar{\bf T}_{-1}$ and  ${\bf F}_{+3}$ are doublets under $S_3$. It belongs to the third row of Table \ref{tab:S4A4}. The $S_4$ tensor product is $\two\otimes\two=\two\oplus \one\oplus \one'$ which becomes $2\cdot \one\oplus \one' \oplus \one''$ under $A_4$. Thus, there are three independent couplings\footnote{Two $\one$'s are counted as the same entry.} which can be of the form  in the $2\times 2$ subspace (due to doublets in the $T_4^0$ twisted sector),
 \dis{
 \begin{pmatrix}0&0&0\\ 0&a&b\\ 0&c&a \end{pmatrix},
 }
 which is general enough to allow the mixing between  top and charm quarks. With  higher dimensional operators \cite{KimRp19}, the 0 entries will be supplied with small numbers and may fulfill the needed $3\times 3$ matrix for the $\Qem=\frac23$ quark matrix.

\bigskip

\noindent
 The above illustration from a compactification model was intended to show a possibility. To study the flavor problem from string compactification, one needs an explicit model locating all the SM fields in the sectors of the compactification as shown in this section.

\section{Conclusion}\label{sec:Conclusion} 

\noindent
We  constructed quark and lepton mass matrices in an anti-SU(5) GUT with a tetrahedral symmetry $A_4$. In the previous paper \cite{FK19}, we showed the hint of the $A_4$ from the PMNS matrix form with one entry being zero. In this paper,  for a convenience of presentation we chose a basis where $\Qem=-\frac13$ quarks and charged leptons are already diagonalised. Then, matter representation $\bar{\bf T}_{-1}$ contains both a quark doublet and a heavy neutrino $N$.  For  $\Qem=+\frac23$ quark masses  $\bar{\bf T}_{-1}$ coupling to ${\bf F}_{+3}$ is used,  and for  neutrino masses the Weinberg operator of $({\bf F}_{+3})^2$ is used through the see-saw of $\bar{\bf T}_{-1}$. In this sense, the quark and neutrino masses are related by the symmetry $A_4$. One notable feature is the anti-SU(5) breaking achieved by the Higgs fields transforming as anti-symmetric representations of SU(5), $\tenb_{-1}^H\oplus \ten_{+1}^H$. This set reduce the rank-5 anti-SU(5) group down to the rank-4 standard model group \smg.   Finally, a string compactification example is presented. As illustrated in this example,   the definite assignments of the SM fields in the twisted sectors are needed to compare with the CKM and PMNS data.

\color{black}

\bigskip

\noindent
  
\acknowledgments{\noindent We have benefitted from comments of S. K. Kang. This work is supported in part  by the National Research Foundation (NRF) grant  NRF-2018R1A2A3074631.}


\end{document}